\documentclass{article}
\usepackage{spconf,amsmath,graphicx}

\usepackage{enumitem}
\setlist{nosep, leftmargin=14pt}

\usepackage{mwe} 

\usepackage{booktabs}
\usepackage{multirow}
\usepackage{pifont}
\usepackage{enumitem}
\usepackage{enumerate}
\usepackage{amsmath}
\usepackage{paralist}
\usepackage{subcaption}
\usepackage{overpic}
\usepackage{url}

\newcommand{\G}{\mathcal{G}}

\newcommand{\ey}{\textbf{\emph{\emph{y}}}}

\newcommand{\eg}{\textbf{\emph{\emph{g}}}}

\newcommand{\ex}{\textbf{\emph{\emph{x}}}}

\newcommand{\eH}{\textbf{\emph{\emph{H}}}}

\newcommand{\emm}{\textbf{\emph{\emph{m}}}}

\newcommand{\BLC}{\textsc{Bloch}}

\newcommand{\proxnet}{\textsc{DeepUnrolling}}
\newcommand{\proxnetB}{\textsc{DeepUnrolling }}

\title{Deep Unrolling for Magnetic Resonance Fingerprinting}
\name{Dongdong Chen$^{1}$, Mike E. Davies$^{1}$, and Mohammad Golbabaee$^{2}$}
\address{$^{1}$ School of Engineering, University of Edinburgh, UK \\
    $^{2}$ Computer Science Department, University of Bath, UK}

\begin{document}
%
\maketitle
\begin{abstract}
Magnetic Resonance Fingerprinting (MRF) has emerged as a promising quantitative MR imaging approach. Deep learning methods have been proposed for MRF and demonstrated improved performance over classical compressed sensing algorithms. However many of these end-to-end models are physics-free, while consistency of the predictions with respect to the physical forward model is crucial for reliably solving inverse problems. To address this, recently \cite{chen2020compressive} proposed a proximal gradient descent framework that directly incorporates the forward acquisition and Bloch dynamic models within an unrolled learning mechanism. However, \cite{chen2020compressive} only evaluated the unrolled model on synthetic data using Cartesian sampling trajectories. In this paper, as a complementary to \cite{chen2020compressive}, we investigate other choices of encoders to build the proximal neural network, and evaluate the deep unrolling algorithm on real accelerated MRF scans with non-Cartesian k-space sampling trajectories.
\end{abstract}
\section{Introduction}
Magnetic resonance fingerprinting (MRF) is an promising technique for MR imaging acquisition and post-processing, which can significantly reduce the acquisition time for quantitative imaging \cite{ma2013magnetic}. Various dictionary matching (DM) and model-based methods have been proposed for tissue quantification in MRF \cite{ma2013magnetic,davies2014compressed}. However, these methods suffer from the enormous storage and computational overhead, and they generally only use the signal pixel to estimate tissue properties and failed to consider the spatial information of the whole image

In order to solve the shortcomings of the model-based MRF methods, many deep learning techniques have been proposed to replace the dictionary matching and use the convolutional layers to exploit spatial context information \cite{golbabaee2019geometry,fang2019deep,chen2019decomposition,Oksuz2019mrf,fang2019rca}. However, these models are trained in an end-to-end fashion, and unlike model-based compressed sensing algorithms \cite{davies2014compressed}, \emph{without} an explicit account for the known physical acquisition model (i.e. the forward operator) and a mechanism for explicitly enforcing measurement consistency. Recently, Chen et al. proposed a deep unrolling framework \cite{chen2020compressive} inspired by the steps of the iterative proximal gradient descent optimisation algorithm. Deep unrolling adopts learnable shared convolutional layers within a data-driven proximal step, meanwhile explicitly incorporating the acquisition model as a non-trainable gradient step in all iterations. However, in  \cite{chen2020compressive} the deep unrolling framework was only evaluated on the synthetic data with gridded/Cartesian k-space subsampling trajectories (using FFT operations), which are less common for fast MRF acquisition. In this paper, as a complementary to \cite{chen2020compressive}, we investigate the choice of encoder network to build the proximal operator and further evaluate the unrolled model on the real-world MRF dataset with  non-Cartesian k-space sampling trajectories and Non-Uniform FFT (NUFFT) operations.

\begin{figure*}[t]
\begin{center}
\includegraphics[width=0.98\linewidth]{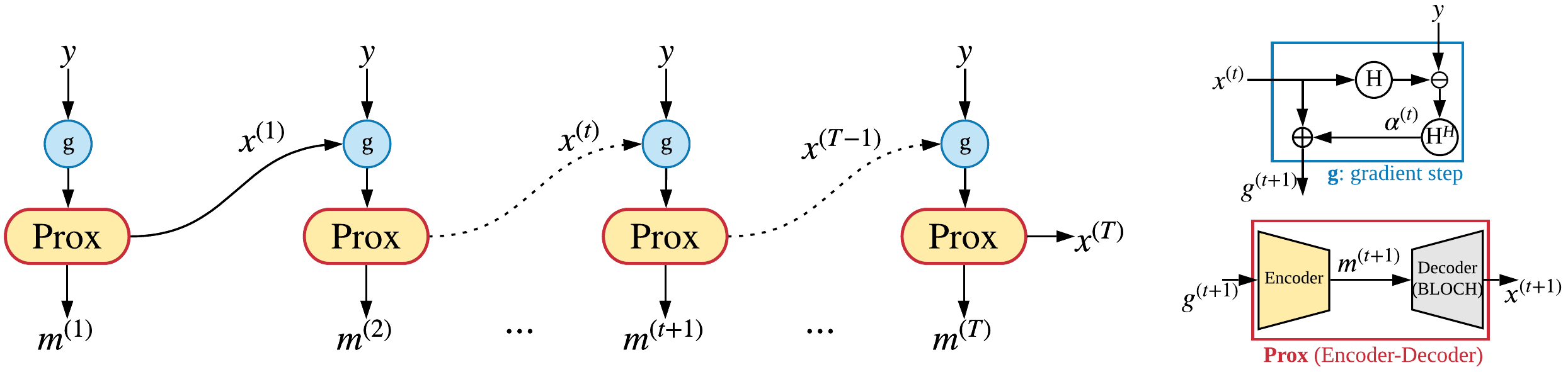}
\end{center}
\caption{Overview of the \proxnet~ for tissue quantification in the compressive MRF.}
\label{fig:unrolling}
\end{figure*}

\section{Deep Unrolling for MRF}

We consider the below MRF inverse problem which adopts a linear spatiotemporal compressive acquisition model:

\begin{equation}\label{eq:sampling2}
    \ey={\eH}({\ex}) +\xi
\end{equation}
where $\ey \in C^{d\times L}$ are the k-space measurements collected at $L$ temporal frames and corrupted by some noise $\xi$.
The acquisition process i.e. the linear forward operator $ \eH: C^{d\times L}\rightarrow C^{D\times s}$ models the multi-coil sensitivity maps,  the (non-uniform) Fourier  subsampling according to a set of temporally-varying k-space locations in each timeframe, combined with a temporal-domain compression scheme for low-rank subspace dimensionality reduction\footnote{ This subspace can be computed through PCA decomposition of the MRF dictionary \cite{mcgivney2014svd,golbabaee2020compressive}} i.e. $s\ll L$. $\ex\in C^{D\times s} $ is the Time-Series of Magnetisation Images (TSMI) across $D$ voxels and $s$ dimension-reduced timeframes (channels).
Accelerated MRF acquisition implies working with under-sampled data which makes the inversion of~\eqref{eq:sampling2} an ill-posed problem.

\noindent\textbf{Bloch response model} Per-voxel TSMI temporal signal evolution is related to the quantitative NMR parameters/properties such as $\{T1_v,T2_v\}$ relaxation times, through the solutions of the \emph{Bloch differential equations}

\begin{equation}
 \overline \ex_v\approx \rho_v \overline{\mathcal{B}}(T1_v,T2_v),
\end{equation}
scaled by the $\rho_v$ proton density (PD) in each voxel $v$~\cite{ma2013magnetic}.

\noindent\textbf{Deep Unrooling.} Typically, the inverse problem (\ref{eq:sampling2}) boils down to solving an model-based first-order iterative \emph{proximal} gradient descent algorithms \cite{davies2014compressed,golbabaee2020compressive}. Recently \cite{chen2020compressive}  proposed to use a neural network to learn the proximal operator (and also the descent step size) from data for MRF reconstruction.
\begin{figure*}[t]
	\centering
	\scalebox{1}{
	\begin{minipage}{\linewidth}
		\centering
\includegraphics[width=1\linewidth]{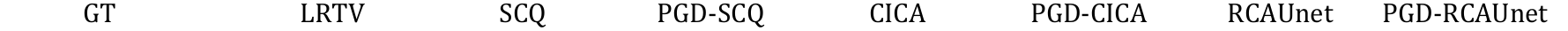}
\\
        \includegraphics[width=.142\linewidth]{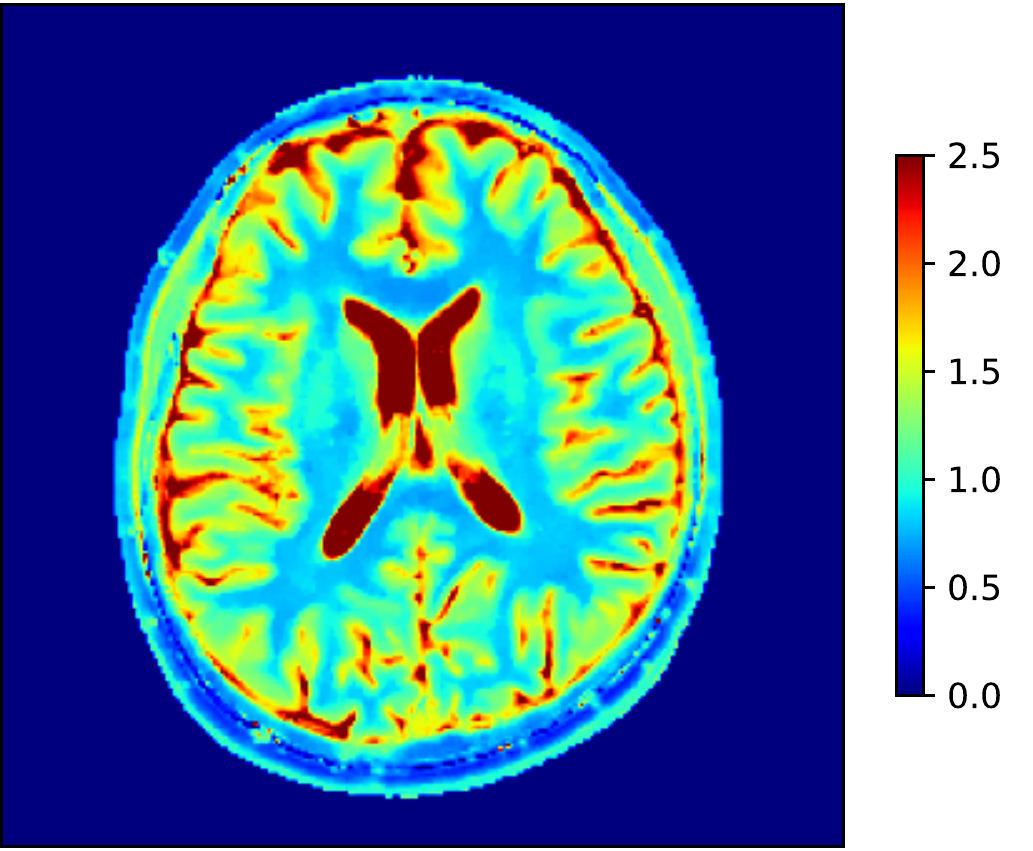}
        \includegraphics[width=.116\linewidth]{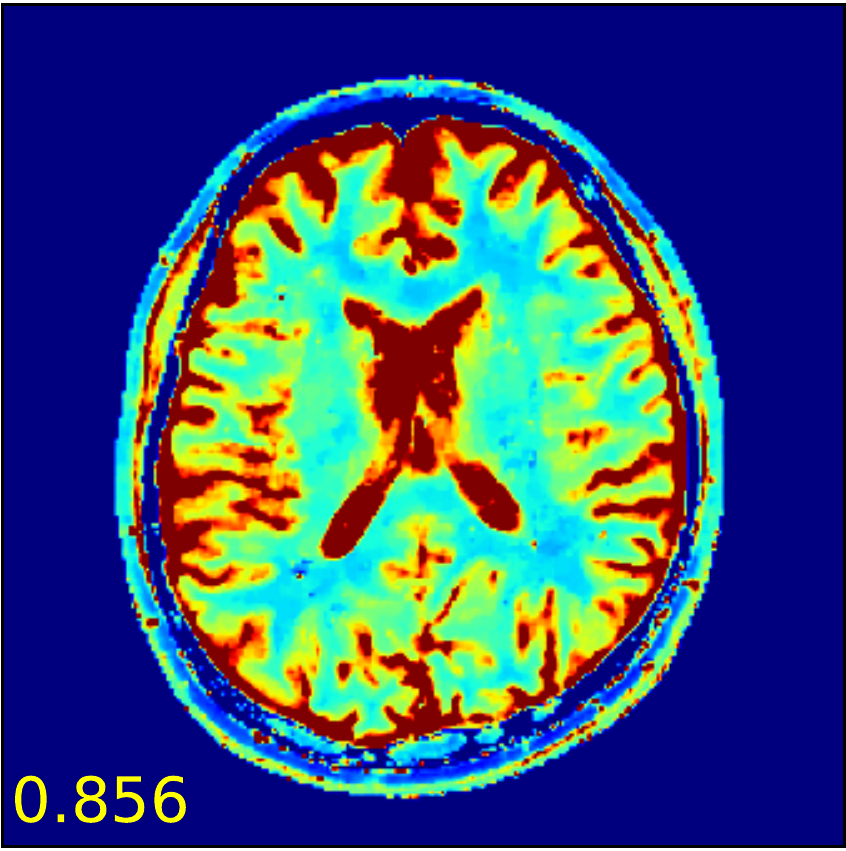}
        \includegraphics[width=.116\linewidth]{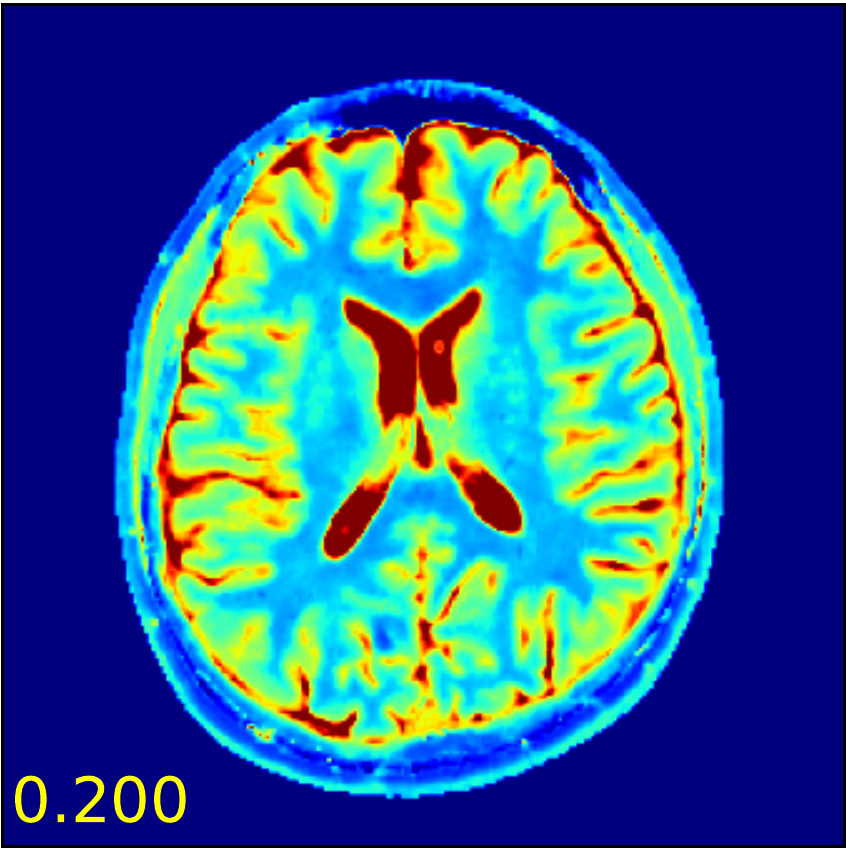}
         \includegraphics[width=.116\linewidth]{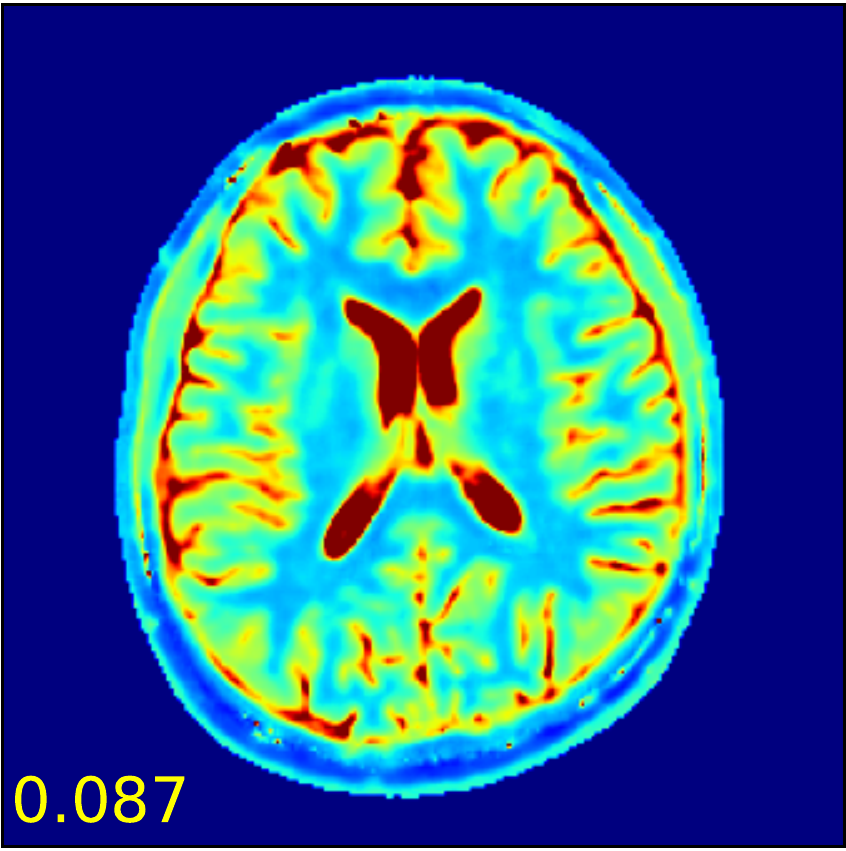}
        \includegraphics[width=.116\linewidth]{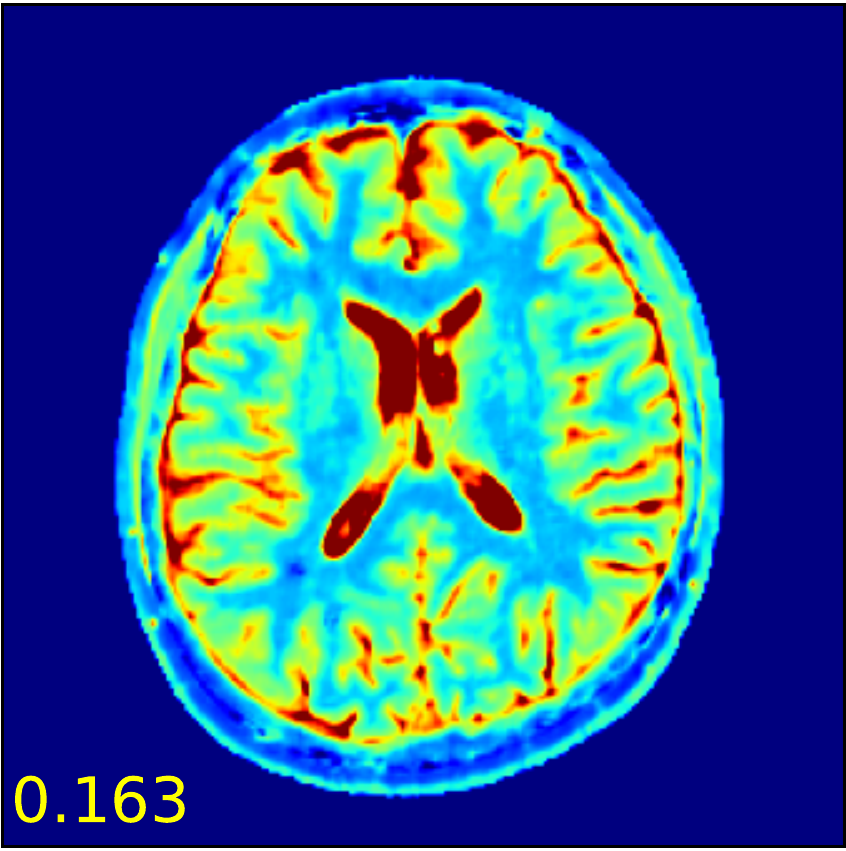}
        \includegraphics[width=.116\linewidth]{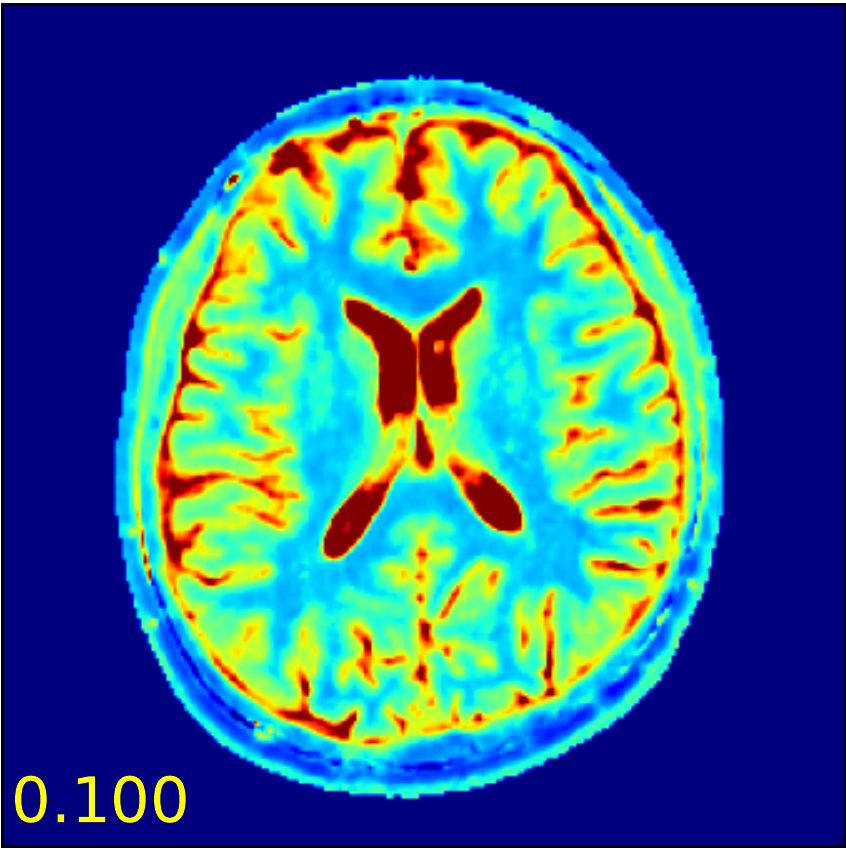}
        \includegraphics[width=.116\linewidth]{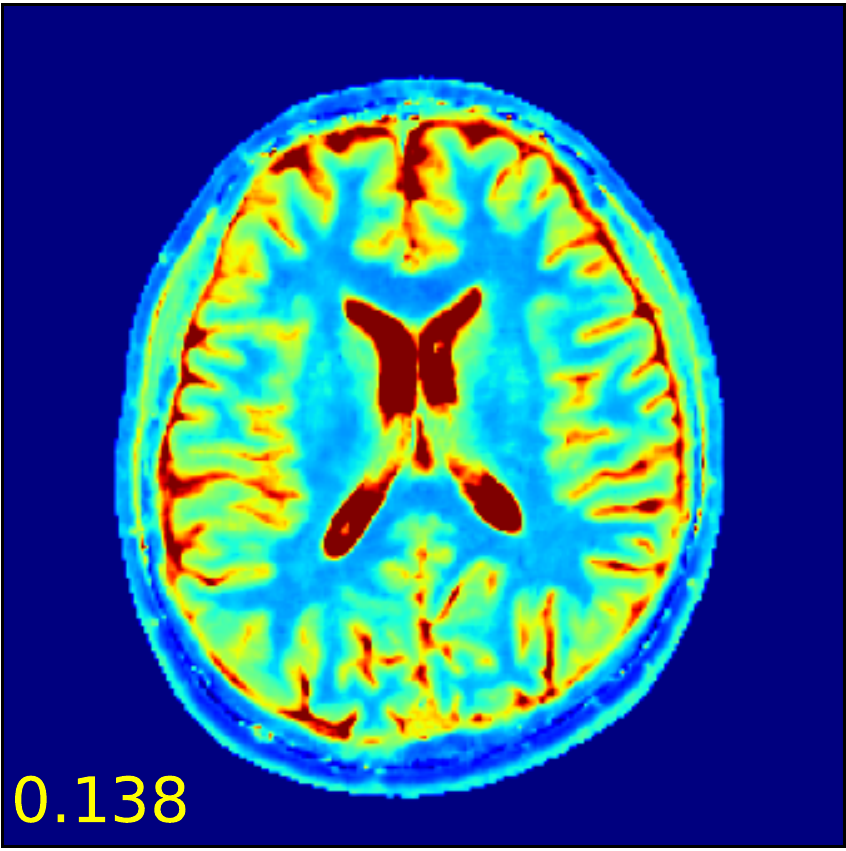}
        \includegraphics[width=.116\linewidth]{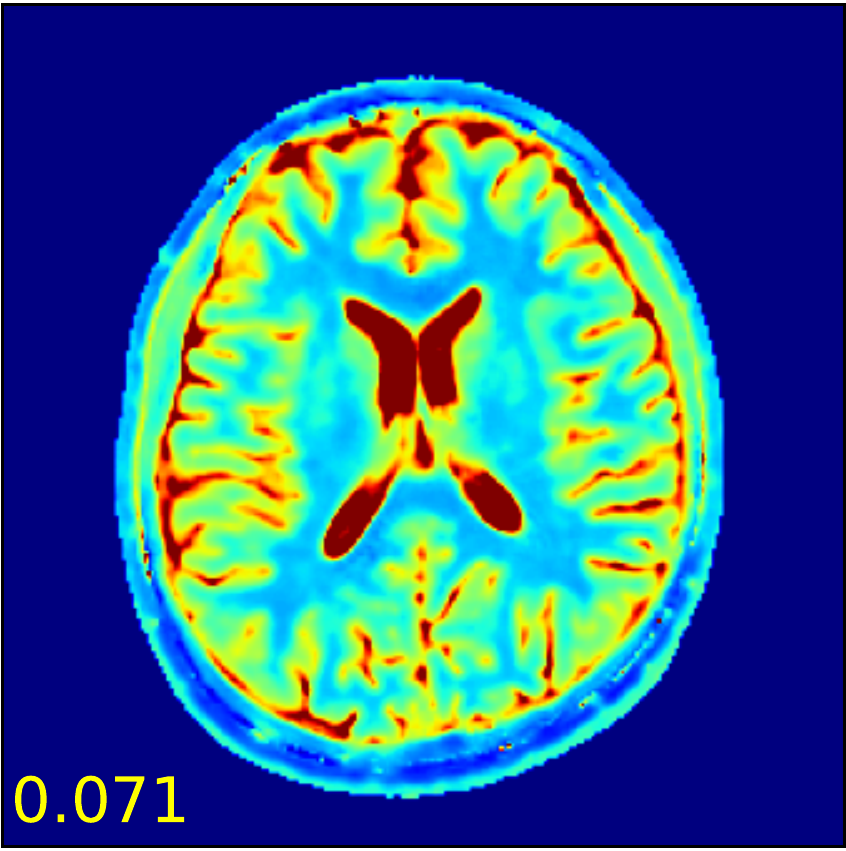}
\\
        \includegraphics[width=.142\linewidth]{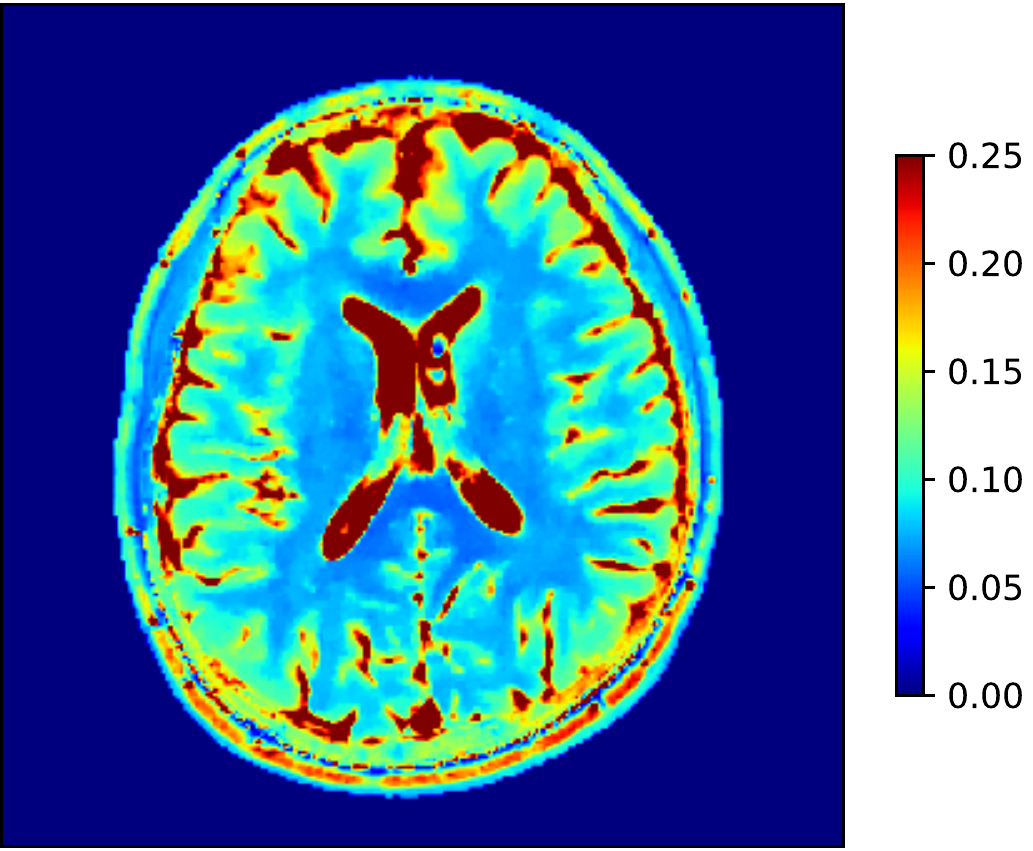}
\includegraphics[width=.116\linewidth]{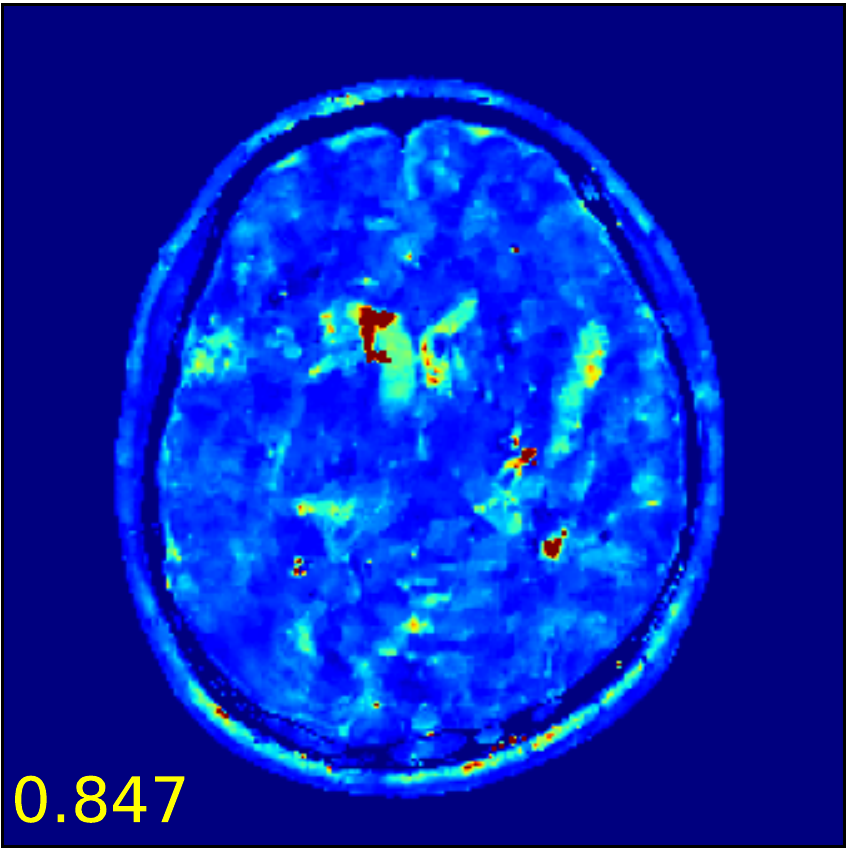}
        \includegraphics[width=.116\linewidth]{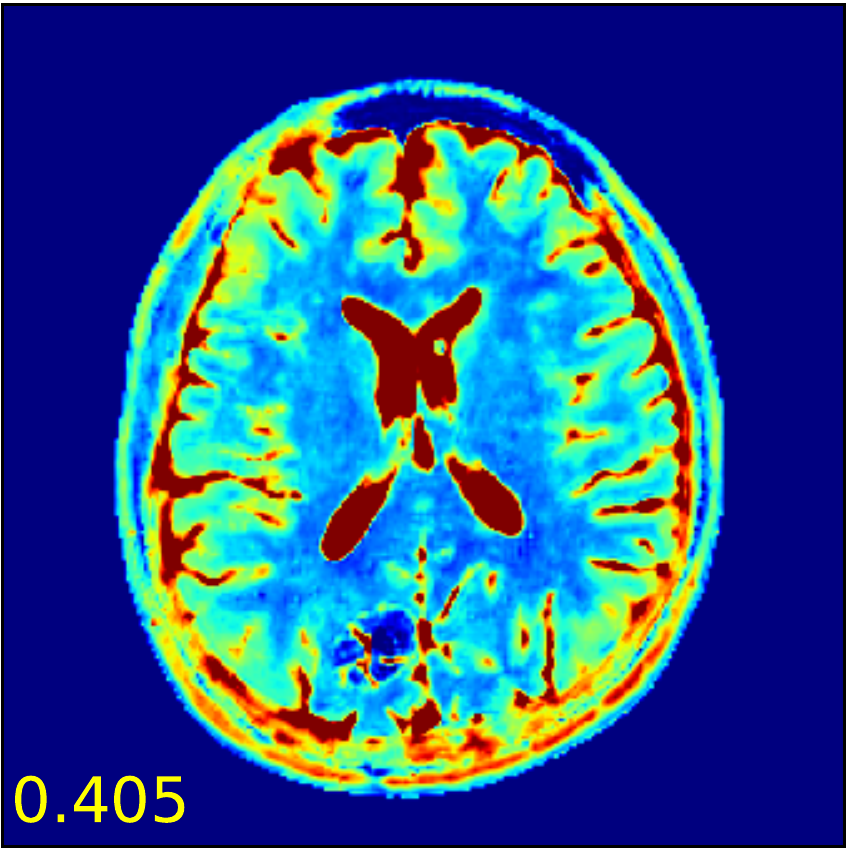}
        \includegraphics[width=.116\linewidth]{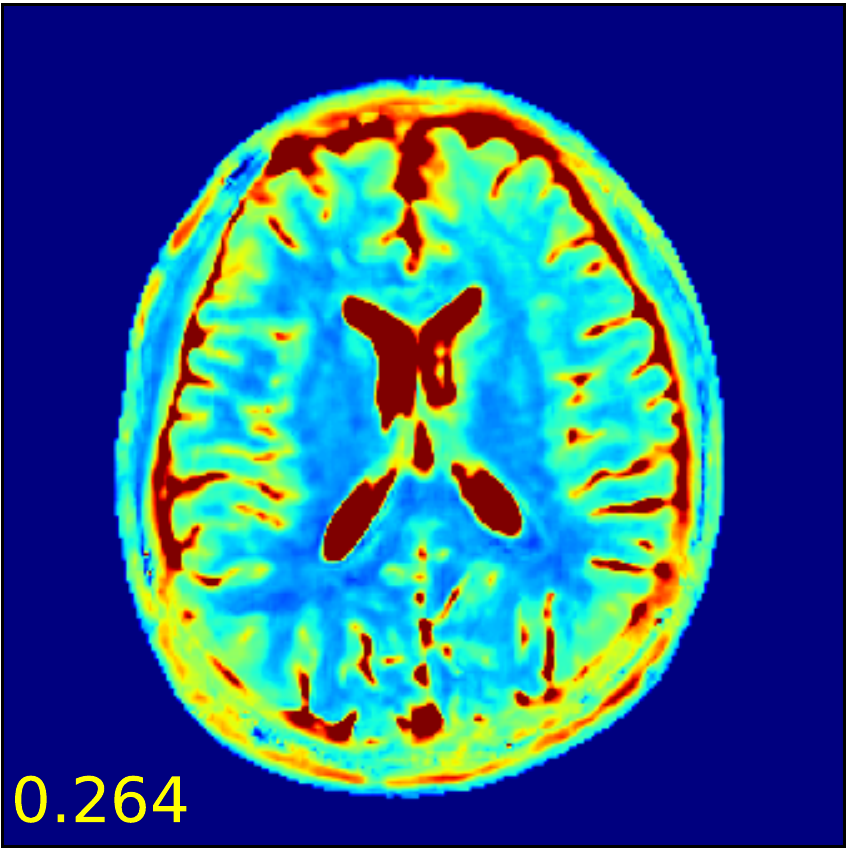}
        \includegraphics[width=.116\linewidth]{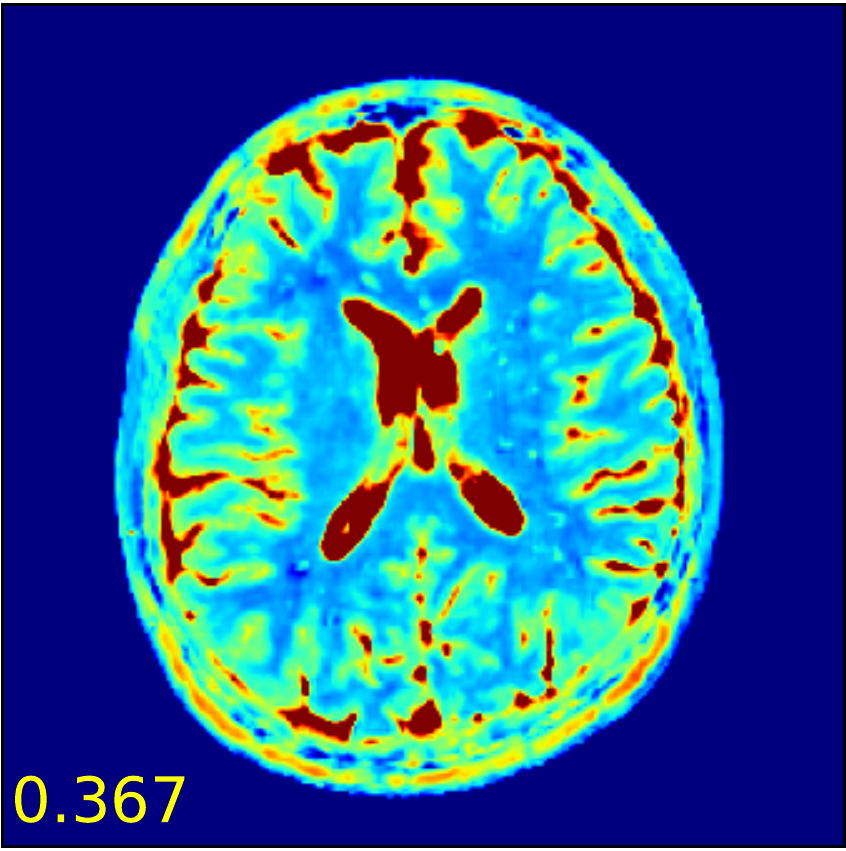}
        \includegraphics[width=.116\linewidth]{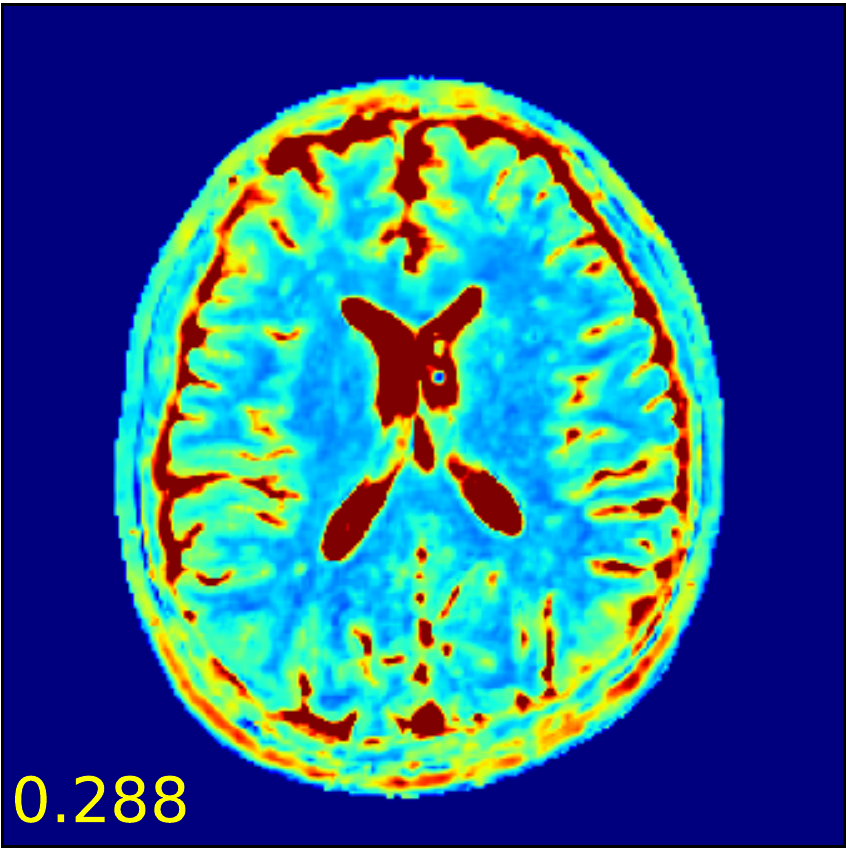}
        \includegraphics[width=.116\linewidth]{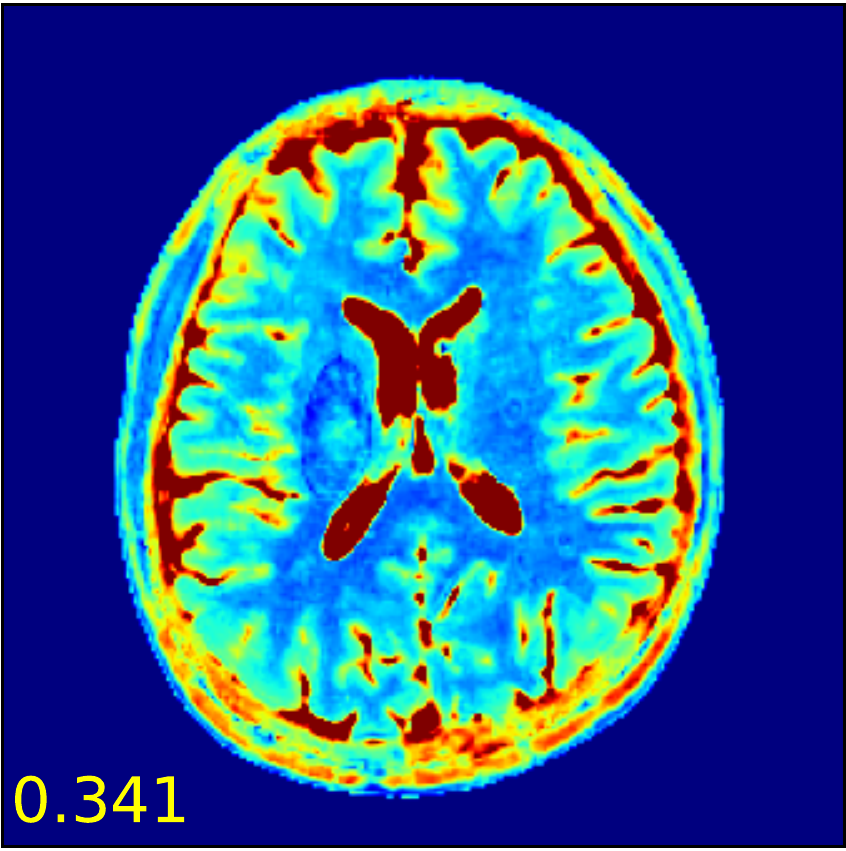}
        \includegraphics[width=.116\linewidth]{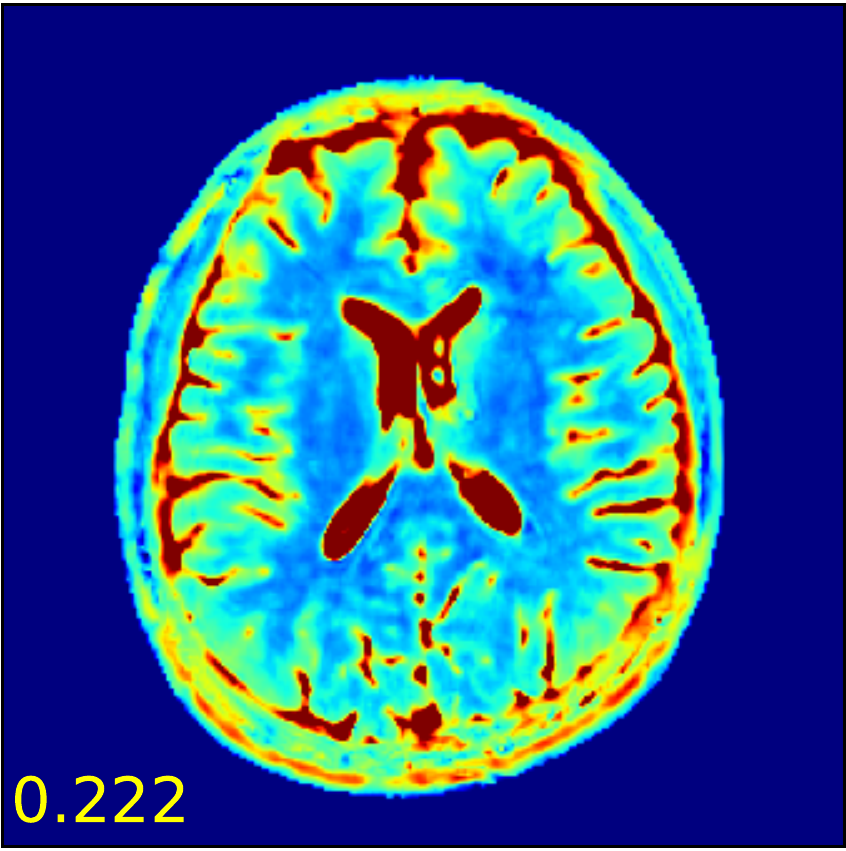}
		\caption{{The reconstruction of T1 map (top) and T2 map (bottom) using LRTV and different encoders (UNet, CICA, RCAUNet and ResNet) with/without using deep unrolling (PGD) calibration on the $10\times$ accelerated data. NRMSE errors are shown in the figures.}  \label{fig:cut3}}
\end{minipage}}
\end{figure*}
Given the compressed measurements $\ey$,
the goal of \proxnetB is to solve the inverse problem~\eqref{eq:sampling2} and to compute the underlying multi-parametric maps $\emm = \{T1, T2, \rho\}$ (and $\ex$ as a bi-product). In particular, as illustrate in Fig.~\ref{fig:unrolling}, \proxnetB aims to solve the below optimization problem:
    \begin{equation}\label{eqs:pnp_regularization}
     \arg\min_{\ex,\emm}\|\ey-\eH\ex\|_2^2 + \phi(\ex,\emm),
    \end{equation}
and solved iteratively by the proximal gradient descent (PGD):
\begin{equation}\label{eq:PDG}
\left\{
\begin{array}{ll}
\eg^{(t+1)} = \ex^{(t)} + \alpha^{(t)}\eH^{H}(\ey- \eH\ex^{(t)}),  & \text{Gradient step}\\ \\
\{\ex^{(t+1)},\emm^{(t+1)}\} = \text{Prox}_{\phi}(\eg^{(t+1)}),& \text{Proximal update}
\end{array}
\right.
\end{equation}
where the gradient updates encourage k-space fidelity (the first term of \eqref{eqs:pnp_regularization}), and the \emph{proximal operator} $\text{Prox}_{\phi}(\cdot)$ enforces image structure priors through a \emph{regularisation} term $\phi(\cdot)$  that makes the inverse problem well-posed.
The Bloch dynamics in \eqref{eq:sampling2} place an important temporal constraint (prior) for per-voxel trajectories of $\ex$. Projecting onto this model (i.e. a temporal Prox model) has been suggested via iterative dictionary search schemes~\cite{davies2014compressed}.

In \cite{chen2020compressive}, the \emph{neural} $\text{Prox}: \eg\rightarrow \{\ex,\emm\}$ is implemented through a deep convolutional \emph{encoder-decoder} network: $\text{Prox}:=\BLC \circ \G$, consisting of an encoder $\G: \eg \rightarrow \emm$ and a decoder \BLC: $\emm \rightarrow \ex$ subnetworks. The \emph{information bottleneck} in the encoder-decoder corresponds to projecting multichannel TSMIs to the low-dimensional manifold of the tissues' intrinsic  (quantitative) property maps~\cite{golbabaee2020compressive}. In particular, the encoder projects  $\eg$ the gradient-updated TSMIs in each iteration (i.e. the first line of \eqref{eq:PDG})  to the quantitative property maps $\emm$. The decoder is a Bloch equation simulation network, creates a differentiable model for \emph{generating} the Bloch magnetic responses.

The target of \proxnet~ framework is to learn a data-driven proximal operator within the PGD mechanism for solving the MRF problem. Implemented by compact networks with convolutional layers, the \emph{neural} Prox improves the storage overhead and the slow runtime of the DM-based PGD by orders of magnitudes. Further, trained on quantitative MR images, the neural Prox network learns to simultaneously enforce spatial- and temporal-domain data structures within PGD iterations.

\noindent\textbf{Training loss.}
Given a training set $\{\emm_i, \ey_i\}_{i=1}^N$, and $T\geq 1$ recurrent iterations of the deep unrolling  (i.e. iterations used in PGD), the loss is defined as
\begin{equation}\label{eqs:loss}
    \min_{\theta,\alpha}\sum\limits_{j\in \Omega}\beta_j \sum\limits_{i=1}^N \ell\left(\emm_{ij},  \emm^{(T)}_{ij}\right) + \lambda \sum\limits_{t=1}^T \sum\limits_{i=1}^N\ell\left(\ey_i, \eH(\ex_i^{(t)})\right),
\end{equation}
where $\Omega=\{T_1, T_2, \rho \}$, $\ell$ is the MSE loss defined with appropriate weights $\beta_j$, $\lambda$ on the tissue property maps $\emm$, and on $\ey$ to maximise k-space data consistency with respect to the forward acquisition model\footnote{Note we removed the reconstruction loss that is defined for $\ex_i$  in \cite{chen2020compressive} since there are no groundtruth $x_i$ for the real dataset.}. In this paper, the scaling between parameters $\beta_j$ and $\lambda$ were initialized based on the physics.

\noindent\textbf{Improvement.} As shown in the Fig.~\ref{fig:unrolling}, the trainable parameters within the \proxnet\ are those of the encoder network $\G$ and the step sizes $\alpha_t$. Other operators such as $\eH, \eH^H$ and \BLC\ (pre-trained separately) are kept frozen during training. Further, $\G$'s parameters are \emph{shared} through all iterations.  In this paper, a truncated $T=5$ PGD iterations are used for training. Supervised training requires the MRF measurements and the ground truth property maps to form the training input $\ey$ and target $ \emm$ samples.
In particular, different from the original \proxnet~\cite{chen2020compressive} where only the synthetic data and the Fast Fourier Transformation (FFT) were used, in this paper, we collect a real MRF dataset and aim to evaluate the model with the real data and in addition, the NUFFT operators that equipped with 8-coil sensitivity complex-valued maps are used. We also evaluated the choice of the encoder network architecture.

\section{Experiments}
The dataset used in this study included 2D axial brain MRF scans of 8 healthy volunteers across 15 slices each.\footnote{Data was obtained from a 3T GE scanner (MR750w system - GE Healthcare, Waukesha, WI) with 8-channel receive-only head RF coil, $230\times230$ mm$^2$ FOV, $230\times230$ image pixels, $5$ mm slice thickness, and used an MRF-FISP acquisition protocol (encoding the T1, T2 and PD properties) with $L=1000$ repetitions, the same flip angles as~\cite{jiang2015mr}, the inversion, repetition and echo times 18, 10, 1.8 (ms) correspondingly.} The ground-truth T1, T2, PD maps were obtained by the LRTV algorithm~\cite{golbabaee2020compressive}.
PCA was applied to obtain $s=10$ channel dimension-reduced TSMI data~\cite{mcgivney2014svd}.
The BLOCH network was pre-trained using the EPG Bloch response simulations \cite{weigel2015extended}.
Data from 7 subjects were used for training our models, and one subject was kept for performance testing.

In this work, we follow \cite{fang2019deep} and (retrospectively) accelerate the MRF acquisition by using fewer time points for tissue quantification. Specifically, for the acceleration rate $r$, we only use the first $\frac{1}{r}\cdot L$ of all $L$ ($L=1000$ in this work) time points. In this paper, we used a challenging $r=10$ acceleration and the number of time points used for tissue quantification is therefore $L_r=100=\frac{1}{10}\times 1000$.

\noindent\textbf{Comparisons.}\label{sec:rec_algos}
We compared against the state-of-the-art MRF baselines LRTV \cite{golbabaee2020compressive}, SCQ \cite{fang2019deep}, RCAUnet \cite{fang2019rca} and CICA \cite{soyak2020channel}. The LRTV is a model-based solution that reconstructs the tissue property maps by solving a low-rank and Total Variation (TV) minimization. SCQ, RCAUnet and CICA are three deep learning solutions,
and both SCQ and RCAUnet are used individual Unets \cite{ronneberger2015u} to separately infer different tissue maps, the RCAUnet and CICA are  channel-attention based methods.
The input to these two networks is the dimension-reduced back-projected TSMIs $\eH^{H}(\ey)$, and their training losses only consider quantitative maps consistency \emph{i.e.} the first term in~\eqref{eqs:loss}.

We trained \proxnet~ with $5$ iterations to learn appropriate encoder $\G$ and the step sizes $\{\alpha^{(t)}\}$. In particular, we applied the CICA, SCQ and RCAUnet as the different encoder networks in the \proxnet~framework. The final hyper-parameters were $\beta=[1, 0.3, 0.6]$ and $\lambda=10^{-3}$ selected via a multiscale grid search to justify their relative weightings in $\beta$ to balance these terms and minimize error w.r.t. the ground truth. The inputs were normalized such that PD ranged in $[0, 1]$.
We used ADAM optimiser with 500 epochs, mini-batch size 1 and learning rate $10^{-3}$. We pre-trained each encoder $\G$ using back-projected TSMIs to initialise the neural network parameters. All algorithms use a $10$-dimensional MRF subspace representation for temporal-domain dimensionality reduction. All networks and operators were implemented in PyTorch and trained and tested on NVIDIA 2080Ti GPUs.

\noindent\textbf{Results and discussion.}
Figure~\ref{fig:cut3} and Table~\ref{tab:testdata1}  compare the performances of the different MRF baselines against their deep unrolled extensions. Reconstruction performances were measured by the NRMSE and MAE.

First, all deep learning methods and their unrolled extensions significantly outperform the model-based LRTV. The unrolled extensions consistently outperform the ordinary ones,  this is achieved due to learning an effective spatiotemporal model (only) for the proximal operator i.e. the $\G$ and \BLC\ networks, directly incorporating the physical acquisition model \eH\ into the recurrent iterations to avoid over-parameterisation of the overall inference model, as well as enforcing reconstructions to be consistent with the Bloch dynamics and the k-space data through the multi-term training loss~\eqref{eqs:loss}. Finally, the RCAUnet performs best in both cases which shows the attention mechanism is helpful for learning better reconstruction. CICA works not as well as RCAUnet though the attention mechanism is also used, one possible reason is that the UNet used in RCAUnet enjoy more helpful inductive bias than that in CICA where only few convolution layers are used.

Compared to popular non-iterative deep learning methods, the proposed deep unrolling method provides a powerful principled framework for constructing interpretable and efficient deep networks. We showed that by a few iterations, unrolling can improve the performance of MRF reconstruction for popular deep learning baselines e.g. CICA, RCAUNet, and SCQ. In our experiments, the unrolled CICA, RCAUNet and SCQ require about 2 seconds to reconstruct a tissue map, while the ordinary (non-iterative) ones only need less than $0.3$ second computation time. Although the recurrent computation usually slows down the training time compared to the non-iterative counterparts, at the inference (testing) time the unrolling method still runs much (2 to 3 orders of magnitude) faster than the conventional model-based reconstruction schemes e.g. LRTV.

\begin{table}[t]
\centering
\fontsize{7}{10}\selectfont
	\caption{Quantitative evaluation (NRMSE and MAE) of the reconstruction quality of the T1, T2 map given by different encoder networks that without (W/O) and with (W) using the PGD unrolling calibration on the $10\times$ accelerated dataset.}
	\vspace{-2pt}
	\begin{tabular}{l|cc|cc}
		\multirow{2}{*}{Encoder}&\multicolumn{2}{c|}{T1} & \multicolumn{2}{c}{T2}\\
		&W/O&W & W/O&W\\
		\toprule
		LRTV\cite{golbabaee2020compressive} & 1.30 (313.13)&-&0.83 (40.62)&-\\
		SCQ\cite{fang2019deep} &0.18 (51.98)&0.09 (33.67)&0.41 (17.63)&0.29 (12.39)\\
		CICA\cite{soyak2020channel} & 0.17 (49.70)& 0.11 (35.18)& 0.37 (15.30)& 0.31 (13.38)\\
		RCAUnet\cite{fang2019rca}& 0.16 (45.27) & 0.07 (26.70)& 0.38 (15.90)& 0.26 (11.25)
 	\end{tabular}
 	\vspace{-10pt}
\label{tab:testdata1}
\end{table}
\section{Conclusions}
Deep unrolling directly incorporates the forward acquisition and Bloch dynamic models within a recurrent learning mechanism with a multi-term training loss.
We validated this approach against a real multi-coil MRF data with non-Cartesian k-space trajectory readouts.
Deep learning outperforms the non data-driven iterative reconstruction algorithm in terms of accuracy and run time.
Through experiments, we observed that several deep learning baselines can be further improved by few iterations of the deep unrolling framework.

\section{Compliance with ethical standards}
\label{sec:ethics}

This research study was conducted retrospectively using anonymised human subject scans made available by GE Healthcare who obtained informed consent in compliance with the German Act on Medical Devices.

\section{Acknowledgments}
\label{sec:acknowledgments}
We thank GE Healthcare for providing the MRF dataset. DC and MD are supported by the ERC Advanced grant C-SENSE, ERC-2015-AdG 694888. MD acknowledges support from his Royal Society Wolfson Research Merit Award.

\bibliographystyle{IEEEtran}
\bibliography{refs}

\end{document}